\documentclass{emulateapj}
\usepackage{natbib}
\newcommand{\msun}{M$_{\sun}$}
\newcommand{\kms}{km s$^{-1}$}
\newcommand{\msuns}{M$_{\sun}~$}

\begin{document}

\title{Binary Capture Rates for Massive Protostars}

\author{Nickolas Moeckel, John Bally} 
\smallskip
\affil{ Center for Astrophysics and Space Astronomy, and\\
        Department of Astrophysical and Planetary Sciences \\
	University of Colorado, Boulder, CO}
\email{moeckel@colorado.edu}

\begin{abstract}
The high multiplicity of massive stars in dense, young clusters is established early in their evolution.  The mechanism behind this remains unresolved.  Recent results suggest that massive protostars may capture companions through disk interactions with much higher efficiency than their solar mass counterparts.  However, this conclusion is based on analytic determinations of capture rates and estimates of the robustness of the resulting binaries.  We present the results of coupled n-body and SPH simulations of star-disk encounters to further test the idea that disk-captured binaries contribute to the observed multiplicity of massive stars.
\end{abstract}


\section{INTRODUCTION}
Massive stars such as those of the Trapezium in the Orion Nebula cluster (ONC) have a high multiplicity compared to low mass stars \citep{mas98,pre99,sta00,gar01}.  The formation of a multiple system in the early life of a cluster can occur through fragmentation of the prestellar material, or by capture.  Capture can occur via multi-body interactions with other cluster members, or through disk interactions in the protostellar phase.  There is growing observational evidence for massive, embedded disks surrounding massive protostars
%, e.g. NGC 7538 IRS 1/2 \citep{kra06}, G192.16-3.82 \citep{she01}, Cepheus A HW2 \citep{pat05}, IRAS 20126+4104 \citep{ces99}, and MWC 349 \citep{whi85,taf04}.  
\citep[][for a recent review]{ces07}.
Simulations of massive star formation suggest that the masses of the disks may build up to values $\sim$ 30\% of the central star's mass before global instabilities trigger a sudden accretion event \citep{kru07}.  Fragmentation of such massive disks into companions is possible \citep{kra06}, but in this work we consider capture of a lower mass cluster member by a massive star-disk system, continuing the analysis presented in \citet{moe06} and \citet[][hereafter MB07]{moe07}.

Disk assisted capture in low mass systems has been studied using analytic methods \citep{cla91} and smoothed particle hydrodynamics (SPH) codes \citep{hel95,bof98}.  The capture rates from these studies are too low to be a significant contributer to the binarity of roughly solar mass stars; for 20 \msuns stars with a 2 \msuns disk in a Trapezium-like environment, MB07 show that the capture rate is high enough to account for 50\% binarity after 1 Myr.  MB07 estimated analytically the likelihood of survival for these binaries, and concluded that higher mass captured companions are more likely to survive, while lower mass companions would be preferentially ionized by encounters with other cluster members.  However, in calculating capture rates and estimating survival fractions, many averages and integrals are taken.  It is desirable to further test these results in an unaveraged, more `realistic' setting.

In this paper we describe the results of n-body simulations of a cluster similar to the ONC, in which we include the dissipative effects of passages through a circumstellar disk surrounding a central 20 \msuns star.  This test demonstrates that the rates derived by MB07 are reasonable in a cluster setting, and that the most likely outcome of encounters between the captured binary and other cluster members leaves the central star in a binary.

\section{SIMULATIONS}
In this study we combine the SPH results of MB07 with n-body simulations of a cluster similar to the ONC.  This is similar in spirit to the work of \citet{sca01} and Pfalzner et al. \citep{pfa06,pfa06a,pfa07}, in which n-body simulations of a cluster are used to determine encounter frequency and parameters, which are then used to study the effect of encounters on disks.  This work is different in that the results of the close encounters are included in the simulation as it is running, similar to \citet{mcd95}, who used an analytic prescription for disk-mediated binary formation in simulations of a cluster with 10 stars. 

The simulations are performed using Aarseth's code NBODY6 \citep{aar00}.  The cluster is set up as a King model \citep[e.g.][]{bin87} with $W_{0}$ = 9, core radius $r_{c}$ = 0.2 pc, central density $n_{0} = 2.0 \times 10^{4}$ pc$^{-3}$, and central velocity dispersion $\sigma_{0}$ = 2.19 \kms.  These parameters are similar to those of the ONC \citep{hil98}, and with a cut-off radius of 2.0 pc yield 4725 stars in the simulation.  The IMF  used is that of \citet{kro02} in the range 0.3 - 9.0 \msuns.  Each star is randomly placed according to the King model density distribution, with a random velocity appropriate to its radius.  In addition, we place a 20 \msuns star at the center of the simulation with a random velocity.  1000 sets of initial conditions were generated; each is run with and without the effects of disk dissipation.  In the dissipation runs, the only star with a disk is the central star.  This is an artificial situation; while the most massive disk will dominate an interaction, the presence of disks around all stars in the cluster would increase the effect of disk encounters.  

MB07 used a modified version of the publicly released SPH code GADGET-2 \citep{spr05} to simulate encounters between the massive star-disk system and impactors with masses in the range 0.3 - 9.0 \msun, periastra in the range 50 - 550 AU, and inclinations from 0$^{\circ}$ - 180$^{\circ}$.  The disk radius is 500 AU.  Interpolation of the data from these simulations gives, for any impactor mass, periastron, and inclination angle, a change in orbital energy and change in disk mass associated with the encounter.  We have modified NBODY6 to detect encounters between the central star and other cluster members.  At the encounter periastron, we determine the orbital parameters, and then modify the velocity of both encounter partners in a momentum conserving fashion so that the change in orbital energy is equal to that found via interpolation of the data from MB07.  The change in disk mass due to the encounter is also tracked.  Because we only have data for periastra $\geq$ 50 AU, encounters with smaller periastra are assumed to occur at 50 AU.

\begin{figure}
 \centering
  \plotone{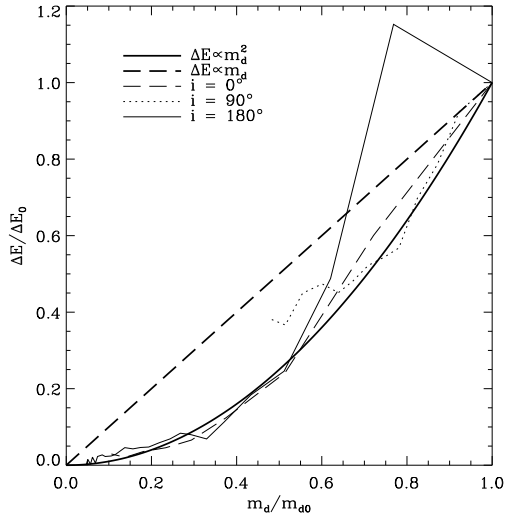}
  \caption{Energy change as a function of remnant disk mass for three of the binaries simulated in \citet{moe06}.  The outlying point for $i$ = 180$^{\circ}$ shows that the true effect of disk disruption on repeated encounters is more complicated than a simple mass scaling.  However, for most encounters, a scaling between the two bold lines is reasonable.}
  \label{energy_scaling}
\end{figure}

When the orbit of a relatively massive impactor is coplanar to the disk, MB07 find that accretion and disk capture can increase the impactor's mass by up to 10\%.  However, the change in orbital energy during these passages is an order of magnitude greater than can be accounted for by accretion drag; the dominant contribution to the energy change is the change in velocity due to disk interactions.  We make the simplification that the change in orbital energy is due entirely to a change in the relative velocity of the two stars.  Under this assumption, at periastron the velocity kick $\delta {\bf v}$ on the impactor for a given orbital energy change $\Delta E$ is given by
\begin{equation}
\label{deltav}
  \delta {\bf v} = \frac{1}{2} \left(-2 v + \sqrt{4v^2 + \frac{8 \Delta E}{m(1+m/M)}} \right) {\bf \hat{v}},
\end{equation}
where $\bf{v}$ is the pre-kick velocity, $m$ is the impactor mass, and $M$ is the primary mass.  The change in velocity for the primary is $\delta {\bf V} = -(m/M) \delta {\bf v}$.  The change in the total energy due to these kicks is tracked and included in energy checks.

In order to account for the change in disk mass, we scale the change in energy for an encounter according to
\begin{equation}
\label{energy_mass_scaling}
  \Delta E(m_{d},i,m_{i},r_{p}) = \Delta E_{0}(i,m_{i},r_{p}) \left( \frac{m_{d}}{m_{d0}} \right) ^{n}.
\end{equation}
Here $\Delta E$ is the orbital energy change from an encounter with disk mass $m_{d}$, impactor mass $m_{i}$, inclination angle $i$, and periastron $r_{p}$.  $\Delta E_{0}$ is the energy change for the same parameters with the disk at its original mass $m_{d0}$, and is found from interpolation of the data in MB07.  The index $n$ we take to be 1 or 2, and scales the proportionality of the the energy change with the remaining disk mass fraction.

\begin{figure}
 \centering
  \plotone{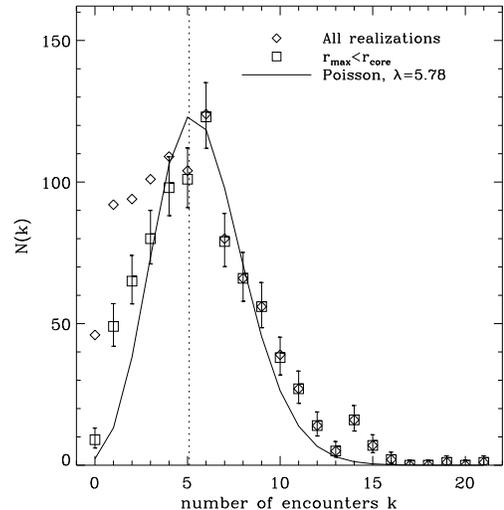}
  \caption{The distribution of encounter frequency for runs with no disk dissipation.  Also plotted are the best fit Poisson distribution to the realizations in which the central star remains in the cluster core ({\it solid}), and the expected value from equation \ref{encounter_rate} ({\it dotted}).  The error bars are single sided 1-$\sigma$ confidence limits \citep{geh86}}
  \label{encounter_distribution}
\end{figure}

The scaling of energy change with disk mass is not completely straightforward.  \citet{hel95}, simulating encounters with disks of different masses, found that the change in energy scales roughly linearly with the disk mass, in which case $n=1$.  \citet{moe06} simulated repeated encounters between a captured impactor and the same remnant disk.  Analysis of that data (Fig. \ref{energy_scaling}) shows that the scaling of energy change and disk mass is more like $n=2$.  However, this also takes into account the changing radius of the disk, an effect which we do not include in this study.  Since the proper scaling depends on details of the encounters that would require full SPH simulations of each close passage, we instead run all simulations with both scalings and compare the results.

Our scheme preserves the binary periastron separation, which is shown by \citet{moe06} to decrease during repeated retrograde encounters, and increase with prograde encounters.  In the most extreme (and rarest) case, an in-plane retrograde passage, the periastron separation decreases by $\sim 25$\% after the first encounter.  The data in MB07 show that the change in energy scales comparably to the periastron radius; thus we would expect an error $\lesssim 25$\% due to our artifically fixed periastron.  This is similar in magnitude to the uncertainty in our mass scaling, which as shown below has a negligible impact on our results.      

Each case is run for 0.5 Myr; we limit the analysis to this time for three reasons.  After this time the disks are mostly destroyed.  In our simulations this destruction is by encounters, while in reality the additional effect of photo-evaporation will contribute.  Mass segregation also begins to take effect after this time, which is an added complication in the analysis of the results.  Finally, the multiplicity of stars is established early in their evolution \citep{mat94}, and a mechanism for binary formation should work on short enough timescales to reflect that.

\section{RESULTS}
\subsection{Encounter Rates}
The calculation of binary capture rates (Clarke \& Pringle 1991; Heller 1995; Boffin et al. 1998; MB07) is largely similar to the estimation of collision or encounter rates.  We begin by comparing the standard encounter timescale calculation to the simulations.

The encounter rate for a star of mass $M$ in a cluster with number density $n$, velocity dispersion $\sigma$, and encounter radius $r$ is given by 
\begin{equation}
\label{encounter_rate}
  \gamma = 4 \sqrt{\pi} n \sigma r^2 \left( 1 + \frac{G {\mathcal M}}{2\sigma^{2}r} \right).
\end{equation}
Here ${\mathcal M}$ is $M + {\bar m}$, with ${\bar m}$ the average stellar mass in the cluster (Binney \& Tremaine 1987; MB07 for general masses).  For the parameters of our simulated clusters, the encounter rate is $\gamma = 1.02 \times 10^{-5}$ yr$^{-1}$, with 5.1 encounters expected over 0.5 Myr.

Plotted in figure \ref{encounter_distribution} is the distribution of the number of unique cluster members encountered by the central star, without disk dissipation.  Multiple encounters with the same star, for instance in a binary, are counted only once.  Because some of the central stars have a high initial velocity and escape to the cluster outskirts, there is an excess of low encounter-number runs.  Therefore the distribution for all runs is plotted, as well as only those in which the central star remained within the cluster core.  One would expect that the distribution for stars in the same environment would be Poisson; the data shows otherwise.  Because the number density and velocity dispersion are radially dependent, central stars that move to larger radii are exposed to a much different environment than those that remain near the cluster center.  By limiting our analysis to the cluster core, where the properties are closest to those used in equation \ref{encounter_rate}, the best fit (with mean value 5.78) is in reasonable agreement with the theoretical value of 5.1.

Equation \ref{encounter_rate} is averaged over the mass function.  Because of the dependence of the encounter rate on the mass of the stars, encounters with more massive cluster members are more frequent.  The mass function of encounter partners is well fit by a single power-law mass function of the form $\xi(m) \propto m^{-\alpha}$ with $\alpha$ = 0.92, while a Salpeter mass function has $\alpha$ = 2.35.

\subsection{Binary Fraction and Mass Function}

Of greatest interest is the fraction of the massive stars at a given time that are in a multiple system.  MB07 calculate that for a cluster with the parameters of our models, the binary formation rate is $\Gamma = 0.6$ Myr$^{-1}$ (Figure 3 in MB07).  We compare this to our simulations as follows.  For each simulation, a list of stars that encounter the central star is generated, with the times of their encounters.  If the same star is involved in consecutive encounters, a binary has formed, and we track the following events.

  {\it Ionization}: The next two encounters are with different stars.  Even if the intruder and the original binary partner form a binary, the central star is no longer in a multiple system.  {\it Exchange}: The next two encounters are with the same star, which is different from the initial binary.  {\it Flyby}: The next encounter is with a different star, but the one after that is with the initial binary partner.  In the latter two cases, the central star is considered to be in a binary throughout.  In this scheme the possibility exists that random, consecutive encounters with the same star could contaminate the binary statistics.  In practice, the number of binaries with semi-major axes larger than the average inter-stellar distance in the cluster core is on the order of 1\%, and we consider the detected binaries to be true binaries. 

\begin{figure}
 \centering
  \plotone{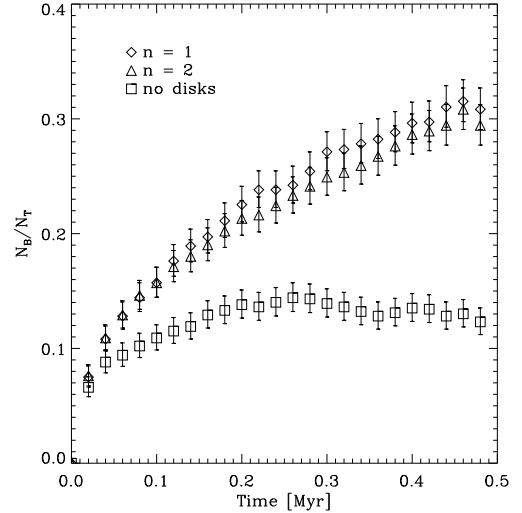}
  \caption{The number of runs with the central star in a binary, $N_{B}(t)$, as a fraction of the total number of runs $N_{T}$ for the three simulation series.  Cases with both $n = 1$ and $n = 2$ in equation \ref{energy_mass_scaling} are at $\sim 30$\% after 0.5 Myr.}
  \label{binary_fraction}
\end{figure}

Plotted in figure \ref{binary_fraction} are the number of central stars in binaries as a function of time, for each of the three series of simulations.  Considering first the series with no disks, we see an initial rise in the binary fraction, followed by a leveling off to $\sim 12$\% at 0.5 Myr, as binary creation through random dynamical processes is balanced by ionization.  For the two series with disk dissipation, there is a steady rise up to a value of approximately 30\%, in good agreement with the calculations in MB07.  The binary fraction does not appear to depend on the specifics of the energy-change disk-mass scaling; the early passages, when the disk still has nearly its original mass, account for the increased binarity.

Since massive impactors are preferentially captured by the disk and more likely to remain bound during exchanges and flybys, the mass function of binary partners at 0.5 Myr is flatter than that of the encounter partners.  With no disks, the best fit single power-law mass function has $\alpha$ = 0.67.  The disk case with $n = 1$ has $\alpha$ = 0.16, and for $n=2$ we have $\alpha$ = 0.12.

\subsection{Binary Robustness}
MB07 estimated the survival probabilities of capture-formed binaries, finding that massive companions are more likely to survive encounters with other cluster members, and that lower mass companions are likely to be ionized.  The binary fractions found here are in agreement with calculations that don't include ionization effects; in order to explain this we turn to the question of ionization versus exchange and flybys.

Shown in table \ref{robustness_table} are $N_{B}$, the number of binaries formed, $N_{E}$, the number of exchanges, $N_{I}$, the number of ionizations, and $N_{F}$, the number of flybys that occur in each series of 1000 simulations.  $N_{B}$ includes all unique binary pairings, so that an exchange contributes twice.  Thus the number of surviving binaries $N_{S}$ is given by $N_{B} - N_{E} - N_{I}$.  The ratio of exchanges and flybys to ionizations, $(N_{E}+N_{F})/N_{I}$, is indicative of the survival chances of a binary in each series.  For the diskless case, $n=1$, and $n=2$ this ratio is 2.12, 8.32, and 6.94 respectively.  Encounters between binaries and other cluster members are much more likely to end with the central star in a binary for the simulations with dissipation compared to the diskless case, and suggest that the binaries formed via this capture mechanism are more robust than indicated by the simple estimations of MB07.  The increased disk dissipation in the $n = 1$ series yields slightly harder binaries, but the total binary fraction is not significantly affected by the scaling.

\begin{deluxetable}{ccccccc} 
\tablecolumns{7} 
\tablewidth{0pc} 
\tablecaption{Binary fate statistics at 0.5 Myr} 
\tablehead{ 
\colhead{Series} & \colhead{$N_{B}$}  &\colhead{$N_{E}$}  &
\colhead{$N_{I}$}  &\colhead{$N_{F}$}   &\colhead{$N_{S}$} &
\colhead{($N_{E}+N_{F})/N_{I}$}
}
\startdata 
no disk & 387 & 44 & 224 & 431 & 119 & 2.12\\ 
n = 1 & 708 & 186 & 217 & 1619 & 305 & 8.32\\ 
n = 2 & 698 & 169 & 238 & 1482 & 291 & 6.94\\ 

\enddata
\label{robustness_table}
\end{deluxetable} 

\section{DISCUSSION}
The simulations presented here are intended to test and verify the capture rates calculated in MB07.  The conclusions of that work are largely upheld; encounters occur at approximately the expected frequency, and binaries are captured at a rate consistent with the analytical estimates.  In addition, once a binary is formed it is less likely to be destroyed by ionization than the estimates in MB07 suggest.  Encounters between the captured binaries and intruding cluster stars are far more likely to result in a flyby or exchange than in an ionization, leaving the central star in a binary.  

The capture rates for the central star-disk system are not high enough to fully account for the high multiplicity observed in massive, cluster-bound stars.  Our simulations produce $\sim 30$\% binarity at 0.5 Myr, the time when the disks are mostly destroyed by encounters or photo-evaporation, an effect not modeled here.  Recent work \citep{kru07} shows that during the formation of a massive star, material moves through the protostellar disk in sporadic, massive accretion events, between which the disk builds up to large masses.  A disk that is $\sim 30-50$\% of the central star's mass, instead of the 10\% used here, could increase the capture rates by a factor of several.  Additionally, continued accretion onto the system or the presence of disks around all the stars could increase the capture rates.  It is worth noting that our simulations here are tailored to ONC-like systems.  Since the capture rate is linear with stellar density and drops with higher velocity dispersion, changing the cluster parameters will affect the results.

The effect of encounters on accretion processes in massive star formation is unclear.  It is possible that the destruction of the disk by repeated passages could truncate accretion at the time of binary formation.  Alternatively, accretion could resume onto the binary and tighten the orbit, leading to a massive, close binary system.  The frequency of the encounters modeled here is high enough that such a situation warrants further investigation.

As concluded in MB07, this rate can not be ignored.  However, additional binary formation mechanisms must be employed to explain the observed multiplicity of massive stars.  In the Trapezium the massive stars have, on average, 1.5 companions \citep{zin02}.  Since disks are effectively destroyed during binary capture, this is a process that can only account for a single companion, unless further accretion creates a new circumstellar or circumbinary disk.

\smallskip

This work was supported by NASA grant NNA04CC11A to the CU Center for Astrobiology.

%%%%%%%%%%%%%%%%%%%%%%%%%%%%%%%%%%%%%%%%%%%%%%%%%%%%%%%%%%%%%%%%%%%%%%%%%%%%%%%%%%%%%%%%%%%%%%%%%%%%%%%%%%%%%%%%%%%%%%%
%%%%%%%%%%%%%%%%%%%%%%%%%%%%%%%%%%%%%%%%%%%%%%%%%%%%%%%%%%%%%%%%%%%%%%%%%%%%%%%%%%%%%%%%%%%%%%%%%%%%%%%%%%%%%%%%%%%%%%%
%%%%%%%%%%%%%%%%%%%%%%%%%%%%%%%%%%%%%%%%%%%%%%%%%%%%%%%%%%%%%%%%%%%%%%%%%%%%%%%%%%%%%%%%%%%%%%%%%%%%%%%%%%%%%%%%%%%%%%%

%\bibliography{masterrefs}

\end{document}